\documentclass{ws-procs9x6}

\def\al{\alpha}
\def\be{\beta}

\def\si{\sigma}

\def\ch{\chi}

\def\om{\omega}

\def\De{\Delta}

\def\Om{\Omega}

\def\to{\rightarrow}

\def\beq{\begin{eqnarray}}
\def\eeq{\end{eqnarray}}

\def\fr#1#2{{{#1} \over {#2}}}
\def\frac#1#2{\textstyle{{{#1} \over {#2}}}}

\def\lsim{\mathrel{\rlap{\lower4pt\hbox{\hskip1pt$\sim$}}
    \raise1pt\hbox{$<$}}}
\def\gsim{\mathrel{\rlap{\lower4pt\hbox{\hskip1pt$\sim$}}
    \raise1pt\hbox{$>$}}}

\def\etal {{\it et al.}}

\def\As#1{({\cal A}_s)_{#1}}
\def\Ac#1{({\cal A}_c)_{#1}}
\def\Bs#1{({\cal B}_s)_{#1}}
\def\Bc#1{({\cal B}_c)_{#1}}
\def\C#1{({\cal C})_{#1}}

\def\nh^#1{{\hat N}^{#1}}
\def\indx{{e\mu}}
\def\indxo{{\bar e\bar\mu}}
\def\nuPOT{6.361\times 10^{20}}
\def\lowEn{544}
\def\oscEn{420}

\newcommand{\refeq}[1]{(\ref{#1})}
\def\etal {{\it et al.}}

\begin{document}

\title{TEST FOR LORENTZ AND CPT VIOLATION WITH 
THE MINIBOONE LOW-ENERGY EXCESS}

\author{T.~KATORI for the MINIBOONE COLLABORATION}

\address{Laboratory for Nuclear Science,
Massachusetts Institute of Technology\\
Cambridge, MA, 02139, USA\\
E-mail: katori@fnal.gov}

\begin{abstract}
The MiniBooNE experiment is a $\nu_\mu\to\nu_e$ and $\bar\nu_\mu\to\bar\nu_e$ 
appearance neutrino oscillation experiment at Fermilab. 
The neutrino mode oscillation analysis shows an excess of 
$\nu_e$ candidate events in the low-energy region. 
These events are analyzed under the SME formalism, 
utilizing the short baseline approximation. 
The preliminary result shows the time independent solution is favored. 
The relationship with the SME parameters extracted from the LSND experiment is discussed. 
The systematic error analysis and antineutrino mode analysis are outlined. 
\end{abstract}

\bodymatter

\section{MiniBooNE low-energy excess}

The main goal of the MiniBooNE experiment\cite{MB-detail}
at Fermilab is to confirm or reject the LSND oscillation signal\cite{LSND}. 
The LSND experiment claimed a $\bar\nu_e$ appearance signal 
from $\bar\nu_\mu$ beam by $\mu^+$ decay at rest 
($\sim$40~MeV), 
and this corresponds to $\De m^2_{LSND}\sim 0.1-1.0$~eV$^2$.   
The MiniBooNE oscillation analysis in neutrino mode rejected this $\De m^2_{LSND}$\cite{MB-osc}$^,$\footnote{
A recent analysis shows an excess for the antineutrino mode\cite{MB-barosc}.}
because MiniBooNE did not see the excess in the energy region 
where the LSND signal is expected ($E_\nu^{QE}>$~475~MeV) 
under two-neutrino massive model. 
However the first oscillation result had unexplained excess at low energy region 
($E_\nu^{QE}<$~475~MeV). 
This was confirmed after a year long reanalysis\cite{MB-unexp}. 
This signal cannot be understood from the three-neutrino massive model, 
but can be understood with a Lorentz violating neutrino oscillation model. 
Especially, a neutrino oscillation model based on Lorentz violation, 
the so called tandem model\cite{tandem}, predicted the low energy excess of MiniBooNE. 
Therefore, it is interesting to search for sidereal variation in 
these low energy excess events.

\section{Data set}

The data set used in this analysis is limited to the neutrino mode. 
Also, all the results presented here are preliminary. 

We used neutrino mode data from March 2003 to January 2006, 
and October 2007 to April 2008, 
corresponding to total of $\nuPOT$ protons on target (POT). 
In this data set, we found $\lowEn$ $\nu_e$ candidate events in the low energy region 
(200~MeV$<E_\nu^{QE}<$475~MeV), 
and $\oscEn$ $\nu_e$ candidate events in the oscillation candidate energy region 
(475~MeV$<E_\nu^{QE}<$1300~MeV). 

Before we proceed to the extraction of Standard-Model Extension (SME) parameters
from our data set, 
we performed statistical tests to quantify any statistically significant time variations. 
We employed Peason's $\ch^2$ (P-$\ch^2$) test and the Kolmogorov-Smirnov (KS) test\cite{Frodesen}. 
The results are shown in Table~\ref{tab:stattest}. 
For the P-$\ch^2$ test, we chose the bin size so that the predicted number of events in each bin is 
greater than 5. The KS test is performed on the unbinned data set to maximize the statistical power. 
There are 4 data sets, the low energy data sample and the oscillation candidate energy data sample, 
each of which include GM time distribution 
and sidereal time distribution parts.  
None of the 4 data sets show statistically significant variation, 
in other words, all data sets are consistent with the flat hypothesis. 
  
\begin{table}
\tbl{A summary of preliminary results from statistical tests on the
sidereal and GM time distributions of the $\nu_e$ candidate data. 
The null hypothesis tests compare
the data with a constant time distribution. 
}
{
\begin{tabular}{@{}ccccc@{}}
\hline
\hline
\multicolumn{5}{c}{null hypothesis tests (preliminary)} \\
\hline
 & \multicolumn{2}{c}{low energy region} &  \multicolumn{2}{c}{oscillation candidate energy region} \\
 & sidereal & GM & sidereal & GM \\
\# of events & \multicolumn{2}{c}{\lowEn} & \multicolumn{2}{c}{\oscEn} \\
\multicolumn{5}{l}{\bf Pearson's $\chi^2$:} \\
$N_\mathrm{bins}$ & 107   & 107   & 83   & 82   \\
$\chi^2$          & 107.6 & 106.0 & 69.6 & 76.2 \\
$P(\chi^2)$       & 0.47  & 0.51  & 0.85 & 0.66 \\
\multicolumn{5}{l}{\bf Kolmogorov-Smirnov:} \\
$P(\mathrm{KS})$  & 0.42  & 0.13  & 0.81 & 0.64 \\
\hline 
\hline
\end{tabular}
\label{tab:stattest}}
\end{table}

\section{Analysis and results}

Although there is no statistically significant time variation in any of the data sets, 
small time variations are not rejected. 
So we continue to search for Lorentz violation in our data sets. 
This analysis follows one performed by LSND collaboration\cite{LSND-LV}. 
We use the SME formalism in the neutrino sector\cite{Matt}. 
The effective Hamiltonian written under the SME formalism is used to write down 
the neutrino oscillation formula, 
with the assumption that the neutrino baseline
is short enough compared to the oscillation length (short baseline approximation). 
Then, we can get the following oscillation formula for 
the $\nu_\mu$ to $\nu_e$ appearance signal\cite{SBA},
\begin{eqnarray}
P_{\nu_\mu\to\nu_e} & \simeq & \fr{L^2}{(\hbar c)^2} |\, \C\indx 
  +\As\indx \sin\om_\oplus T_\oplus
  +\Ac\indx \cos\om_\oplus T_\oplus  \nonumber \\
 & &
  +\Bs\indx \sin2\om_\oplus T_\oplus
  +\Bc\indx \cos2\om_\oplus T_\oplus \, |^2~.
\label{eqn:prob}
\end{eqnarray}
Here $L$ is the baseline of neutrinos, 
$\Om_\oplus$ is the sidereal frequency (=2$\pi$/23h 56min 4.1s), 
and $T_\oplus$ is the sidereal time. 
The 5 parameters $\C\indx$, $\As\indx$, $\Ac\indx$, $\Bs\indx$, and
$\Bc\indx$ depend on the SME coefficients $(a_L)^{\al}_{\indx}$ and 
$(c_L)^{\al\be}_{\indx}$\cite{SBA}. 

We use the unbinned likelihood method to find these parameters. 
However, in these proceedings, 
fitting parameters are limited to the first 3 parameters ($\C\indx$, $\As\indx$, and $\Ac\indx$). 
This 3 parameter fit can be interpreted as the case where Nature 
only has CPT-odd Lorentz violating coefficients. 

\begin{figure}[ht]
\centerline{\epsfxsize=5.0in\epsfbox{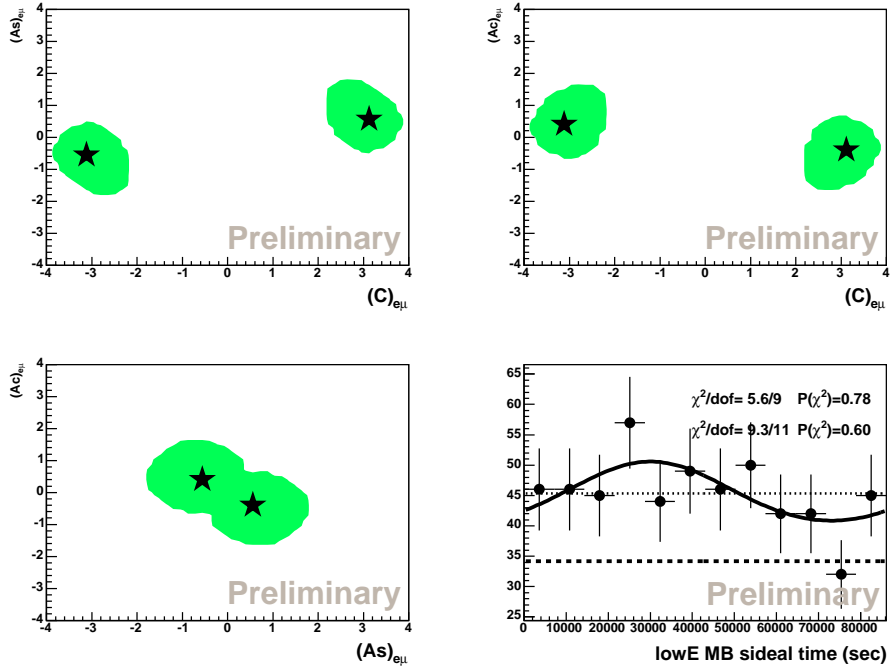}}   
\caption{The preliminary unbinned likelihood fit 
solution for the low energy region 
sidereal time distribution. 
The first 3 plots show parameter space 
with the preliminary best fit point and 1$\si$ contours, 
and the last plot shows the preliminary best fit solution 
and its goodness-of-fit. 
The curve shows the model line with the preliminary best fit parameters, 
whereas dotted and dashed lines show 
the flat solution and background. 
The upper $\ch^2$ and $P(\ch^2)$ are after the fit, 
the lower ones are before it (flat solution goodness-of-fit). 
\label{fig:lowE}}
\end{figure}

The fitting results for the low energy region data samples are shown in Fig.\ \ref{fig:lowE}. 
The solutions are duplicated due to the nature of Eq.~\refeq{eqn:prob}. 
From Fig.~\ref{fig:lowE}, one can see that only $\C\indx$ (sidereal time independent term) 
differs from zero with statistical significance. 
This means that the data favor the flat solution, 
and this is consistent with the flatness tests in previous section. 
Also, the increase of goodness-of-fit after the unbinned likelihood fit is small 
($P(\ch^2)=0.60\to P(\ch^2)=0.77$). 
This is expected because the flat hypothesis already provides a very good fit. 
The situation is slightly different 
for the oscillation candidate energy region. 
Since the excess itself is small, 
all parameters are statistically consistent with zero.

Finally, the SME coefficients with 1$\si$ statistical errors 
are extrapolated from the fit results. 
Note, 350~MeV is used as an averaged neutrino energy 
for the low energy region data sample.  
From $\C\indx$, an extrapolation is performed 
assuming only one SME coefficient is nonzero. 
$\As\indx$ and $\Ac\indx$ are solved assuming two coefficients are nonzero. 
The results are summarized in Table~\ref{tab:fitresult}. 
We also calculate SME coefficients from the LSND experiment\cite{LSND-LV}. 
The 3 parameter fit result from LSND has 2 solutions in 1$\si$ area, 
but we accept the statistically significant solution. 
We used 40~MeV for the averaged neutrino energy.

\begin{table}
\tbl{The preliminary extrapolated SME coefficients 
from the 3 parameter fit result. 
The 3 missing rows corresponds to $\Bs\indx$ and $\Bc\indx$, 
which we did not use for the fit at this time. 
We use 350~MeV (MiniBooNE) and 40~MeV (LSND) 
for the averaged neutrino energies for these  extrapolations. 
The numerical entries are preliminary extracted values with 1$\si$ statistical errors.}
{
\begin{tabular}{@{}cccc@{}}
\hline
\hline
\\[-8pt]
\multicolumn{4}{c}{Preliminary SME coefficients 
from the 3 parameter unbinned likelihood fit} \\
\hline
\\[-8pt]
SME &result from&SME &result from\\
coefficient &MiniBooNE  &coefficient&LSND\\
\hline
\\[-5pt]
$(a_L)^T_{\indx}$   &$(-3.1\pm 0.9)\times 10^{-20}$~GeV&$(a_L)^T_{\indxo}$   &$ (0.2\pm 1.0)\times 10^{-19}$~GeV\\
$(a_L)^X_{\indx}$   &$ (0.6\pm 1.9)\times 10^{-20}$~GeV&$(a_L)^X_{\indxo}$   &$ (4.2\pm 1.5)\times 10^{-19}$~GeV\\
$(a_L)^Y_{\indx}$   &$(-0.9\pm 1.8)\times 10^{-20}$~GeV&$(a_L)^Y_{\indxo}$   &$(-1.7\pm 1.8)\times 10^{-19}$~GeV\\
$(a_L)^Z_{\indx}$   &$(-4.2\pm 1.2)\times 10^{-20}$~GeV&$(a_L)^Z_{\indxo}$   &$ (1.0\pm 5.4)\times 10^{-19}$~GeV\\
$(c_L)^{TT}_{\indx}$&$ (7.2\pm 2.1)\times 10^{-20}$    &$(c_L)^{TT}_{\indxo}$&$ (0.3\pm 1.8)\times 10^{-18}$\\
$(c_L)^{TX}_{\indx}$&$(-0.9\pm 2.8)\times 10^{-20}$    &$(c_L)^{TX}_{\indxo}$&$(-5.2\pm 1.9)\times 10^{-18}$\\
$(c_L)^{TY}_{\indx}$&$ (1.3\pm 2.6)\times 10^{-20}$    &$(c_L)^{TY}_{\indxo}$&$ (2.1\pm 2.2)\times 10^{-18}$\\
$(c_L)^{TZ}_{\indx}$&$ (5.9\pm 1.7)\times 10^{-20}$    &$(c_L)^{TZ}_{\indxo}$&$ (1.3\pm 6.7)\times 10^{-18}$\\
$(c_L)^{XX}_{\indx}$&               ---                &$(c_L)^{XX}_{\indxo}$&             ---             \\
$(c_L)^{XY}_{\indx}$&               ---                &$(c_L)^{XY}_{\indxo}$&             ---             \\
$(c_L)^{XZ}_{\indx}$&$(-1.1\pm 3.7)\times 10^{-20}$    &$(c_L)^{XZ}_{\indxo}$&$(-2.7\pm 1.0)\times 10^{-17}$\\
$(c_L)^{YY}_{\indx}$&               ---                &$(c_L)^{YY}_{\indxo}$&             ---             \\
$(c_L)^{YZ}_{\indx}$&$ (1.7\pm 3.4)\times 10^{-20}$    &$(c_L)^{YZ}_{\indxo}$&$ (1.1\pm 1.2)\times 10^{-17}$\\
$(c_L)^{ZZ}_{\indx}$&$ (2.6\pm 0.8)\times 10^{-19}$    &$(c_L)^{ZZ}_{\indxo}$&$(-1.1\pm 5.9)\times 10^{-18}$\\
\\[-5pt]
\hline
\hline
\multicolumn{4}{c}{}
\end{tabular}
\label{tab:fitresult}}
\end{table}

For the MiniBooNE result, we can see $(a_L)^T_{\indx}$, $(a_L)^Z_{\indx}$, 
$(c_L)^{TT}_{\indx}$, $(c_L)^{TZ}_{\indx}$, 
and $(c_L)^{ZZ}_{\indx}$ are statistically significant from zero. 
This is the consequence of $\C\indx$ being nonzero from 
the 3 parameter fit result. 
On the other hand, since $\As\indxo$ is nonzero for LSND, 
$(a_L)^X_{\indxo}$, $(c_L)^{TX}_{\indxo}$, and $(c_L)^{XZ}_{\indxo}$ 
are statistically significant. 
Since the MiniBooNE result provides neutrino SME coefficients, 
and the LSND result provides antineutrino SME coefficients, 
these two results are not inconsistent. 

In conclusion, we performed a search for Lorentz violation 
in the MiniBooNE low energy excess sample. 
We discovered that the time independent solution is favored. 
To complete the analysis, we are planning a study of systematic errors. 
We also plan to extend this analysis to the antineutrino mode. 
This is very interesting because the antineutrino mode has an excess 
in the oscillation candidate energy region\cite{MB-barosc}. 

\section*{Acknowledgments}

We thank Ben Jones for a careful reading of this manuscript.


\begin{thebibliography}{xx}

\bibitem{MB-detail} 
A.~A.~Aguilar-Arevalo \etal, MiniBooNE Collaboration, 
Phys.~Rev.~D {\bf 79}, 072002 (2009); 
Nucl.~Instr.~Meth.~A {\bf 599}, 28 (2009). 

\bibitem{LSND}
A.~A.~Aguilar-Arevalo \etal, LSND Collaboration,
Phys.~Rev.~D {\bf 64}, 112007 (2001).

\bibitem{MB-osc}
A.~A.~Aguilar-Arevalo \etal, MiniBooNE Collaboration,
Phys.~Rev.~Lett. {\bf 98}, 231801 (2007).

\bibitem{MB-barosc}
A.~A.~Aguilar-Arevalo \etal, MiniBooNE Collaboration, 
arXiv:1007.1150v1; 
R. Tayloe, these proceedings.

\bibitem{MB-unexp}
A.~A.~Aguilar-Arevalo \etal, MiniBooNE Collaboration,
Phys.~Rev.~Lett. {\bf 102}, 101802 (2009).

\bibitem{tandem}
T.~Katori,~R. Tayloe, and V.A.~Kosteleck\'{y},
Phys.~Rev.~D {\bf 74}, 105009 (2006).

\bibitem{Frodesen} 
A. Frodesen, O. Skjeggestad, and H. T\o fte, 
{\em Probability and Statistics in Particle Physics}, 
Universitetsforlaget, Bergen, 1979.

\bibitem{LSND-LV}
L.B.~Auerbach \etal, LSND Collaboration,
Phys.~Rev.~D {\bf 72}, 076004 (2005).

\bibitem{Matt}
V.A.~Kosteleck\'{y} and M. Mewes,
Phys.~Rev.~D {\bf 69}, 016005 (2004).

\bibitem{SBA}
V.A. Kosteleck\'{y} and M. Mewes,
Phys. Rev. D {\bf 70}, 076002 (2004).

\end{thebibliography}
\end{document}